\documentclass[ reprint,
 amsmath,amssymb,
 aps,
]{revtex4-1}

\usepackage{color}
\usepackage{graphicx}
\usepackage{dcolumn}
\usepackage{amsbsy}
\usepackage{bm}

\newcommand{\be}{\begin{equation}}
\newcommand{\ee}{\end{equation}}
\newcommand{\bea}{\begin{eqnarray}}
\newcommand{\eea}{\end{eqnarray}}

\newcommand{\s}{\sigma}
\newcommand{\la}{\langle}
\newcommand{\ra}{\rangle}

\newcommand{\lp}{\left(}
\newcommand{\rp}{\right)}
\newcommand{\Tr}{{\rm Tr\,}}

\newcommand{\<}{\langle}
\renewcommand{\>}{\rangle}

\renewcommand{\Im}{{\rm Im\,}}
\renewcommand{\vec}[1]{{\bf #1}}
\renewcommand{\hat}[1]{{\widehat #1}}
\renewcommand{\phi}{\varphi}
\renewcommand{\epsilon}{\varepsilon}

\newcommand{\ve}{\varepsilon}
\newcommand{\sgn}{\text{sign}}

\begin{document}

\preprint{}

\title{Robustness and universality of surface states in Dirac materials}

\author{Oles Shtanko, Leonid Levitov}
\affiliation{Department of Physics, Massachusetts Institute of Technology, 77 Massachusetts Avenue, Cambridge, Massachusetts 02139, USA}

\begin{abstract}
Ballistically propagating topologically protected states harbor
exotic transport phenomena of wide interest. Here we describe
a nontopological mechanism that produces such states at the
surfaces of generic Dirac materials, giving rise to propagating
surface modes with energies near the bulk band crossing. The
robustness of surface states originates from the unique properties of Dirac-Bloch wavefunctions which exhibit strong coupling
to generic boundaries. Surface states, described by Jackiw-Rebbi-type bound states, feature a number of interesting properties.
Mode dispersion is gate tunable, exhibiting a wide variety of
different regimes, including nondispersing flat bands and linear
crossings within the bulk bandgap. The ballistic wavelike character of these states resembles the properties of topologically
protected states; however, it requires neither topological restrictions nor additional crystal symmetries. The Dirac surface states
are weakly sensitive to surface disorder and can dominate edge
transport at the energies near the Dirac point.
\end{abstract}

\maketitle

Surface states and the mechanisms allowing them to propa-
gate along crystal boundaries -- the topics of long-standing
interest of the theory of solids -- acquired a new dimension with
the advent of topological materials
\cite{hasan2010colloquium,wehling2014dirac}. 
In these materials
robust surface states are made possible by nontrivial topology
of the bulk bands \cite{kane2005z,hasan2010colloquium}. 
Here we outline a different mechanism leading to robust surface states, realized in solids with
Dirac bands that mimic relativistic particles near band crossings \cite{wehling2014dirac}. 
In this scenario robust surface states originate from
unusual scattering properties of Dirac particles, occurring for
generic boundary conditions at the crystal boundary. As we will
see, since this mechanism does not rely on band topology, it
can lead to robust surface states in solids with either topological or nontopological bulk band dispersion. The surface states
exist for either gapless or narrow-gapped Dirac bulk bands. Furthermore, these states are to some degree immune to surface
disorder. Namely, as discussed below, surface modes can propagate coherently by diffracting around surface disorder through
system bulk. This diffraction behavior suppresses backscattering
and results in exceptionally long mean free paths. Since Dirac
surface states require neither special topological properties of
the band structure nor special symmetry, they are more generic
than the topological surface states. As such, these states can shed
light on recent observations of edge transport in nontopological
materials.

Indeed, it is often taken for granted that an observation of edge transport signals nontrivial band topology \cite{Konig_2007,Roth_2009,Nowack_2013,Pribiag_2015}.  
However, recent experiments on semiconducting structures, where tunable band inversion enables switching between topological and non-topological phases, indicate that current-carrying edge modes can appear regardless of the band topology \cite{Nichele_2016,Knez_2014,Du_2015,Ma_2015}. 
One piece of evidence comes from transport and scanning measurements in InAs/GaSb, which indicate that helical edge channels survive switching from a topological to a trivial band structure \cite{Nichele_2016}.
Additionally, Refs. \cite{Knez_2014} and \cite{Du_2015} report an unexpectedly weak dependence of edge transport on the in-plane magnetic field. Namely, it is found that the edge transport is observed even when Zeeman splitting is considerably larger than the spin-orbit splitting, i.e., in the nontopological regime. A similar behavior
is observed in HgTe devices \cite{Ma_2015}. 
Furthermore, recently several
groups have used Josephson interferometry to directly image
long-range edge currents in graphene, a signature nontopological material \cite{allen2016spatially,allen2015visualization,shalom2016quantum,calado2015ballistic,zhu2017edge}. 
These observations point to the existence of robust
nontopological surface states.

\begin{figure}[t] 
\center{\includegraphics[width=1\linewidth]{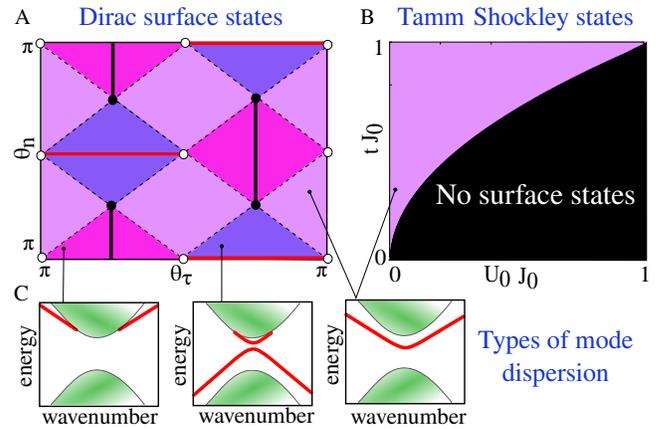}}
\caption{
(\textit{A} and \textit{B}) Phase diagrams for Dirac surface states (\textit{A}) and Tamm-Shockley surface states (\textit{B}) as a function of parameters that control boundary conditions (main text). Dirac surface states occur in the entire parameter
space except for a subset of measure zero (black lines) and are therefore generic. In contrast, the Tamm-Shockley surface states appear upon fine tuning, occurring in a relatively small region of the parameter space,
whereas most of phase space is incompatible with surface states. Different types of Dirac surface bands, shown in \textit{C}, correspond to regions of different color (purple, blue, and pink) in \textit{A} and \textit{B}. For a detailed discussion see
main text.
}
\label{fig:phase_diagram}
\end{figure}

As we will see, the Dirac surface states can arise naturally
due to strong coupling of electronic waves to generic boundaries. Namely, the phase shifts of waves in the bulk that scatter off the surface have a strong energy dependence near the Dirac
point where the particle and “antiparticle” bands cross (or nearly
cross). The energy dependence of phase shifts, as always, leads
to the formation of states behaving as plane waves confined to
the surface and decaying into the material bulk as evanescent waves. The formation of these states is governed by a mechanism that resembles the seminal Jackiw–Rebbi (JR) theory \cite{jackiw1976solitons}
for the states formed at the domain walls separating regions with sign-changing Dirac mass. Unlike the JR problem, however, the Dirac surface states do not have a topological character i.e. in general they are not protected by topological invariants. Unlike the JR problem, however, the
Dirac surface states do not have a topological character; i.e., in general they are not protected by topological invariants. Nevertheless, these states are robust and form surface modes with the energies near the Dirac crossing of bulk bands (see Fig.1C). 

The diffraction-based mechanism that suppresses backscattering and makes the Dirac surface states insensitive to surface
disorder has an interesting analogy with the properties of the high-mobility electron gas realized in GaAs/AlAs quantum wells.
In these systems an exceptionally high mobility could be achieved
by adjusting the well width to reduce the overlap of the carrier scattering at the surface disorder. 
Scattering suppression through
this mechanism results in a dramatic increase of the mean free path, growing rapidly vs. the well width, $\ell\sim w^n$ with large $n$ \cite{gold1986metal}. 
Recently in wide quantum wells mobilities exceeding $10^7 \,{\rm cm^2/V\, s}$ 
were demonstrated \cite{kamburov2016interplay}.  
Likewise, Dirac surface
states, being nontopological, are, in principle, susceptible to disorder. However, the large width of Dirac surface states (arising
due to their slow decay into the bulk) can strongly reduce their
overlap with the atomic-scale disorder at the surface and make
them effectively immune to surface scattering. In this regime,
in direct analogy with the carriers in wide quantum wells, the
surface states can propagate coherently by diffracting around
surface disorder. This remarkable behavior is discussed in detail
below. We will see that, while Dirac surface states may be gapped
(Fig. 1) and are generally not immune to the bulk disorder, their
decoupling from the surface disorder can suppress backscattering and enable large mean free paths already for moderately
clean materials. This property also weakens the dependence of
these states on the details of the surface structure.

At this point it is instructive to compare Dirac surface states to the well-known Tamm-Shockley states. These are nontopological states residing inside the bandgap that governs surface physics
of many semiconductors. The Tamm–Shockley states form a surface band that splits off the bandgap edge upon varying the surface potential. The existence of these states depends on the details of the crystal structure near the surface, which makes them nonuniversal and less robust than the Dirac surface states (see \textit{Supplementary Information: The Tamm-Shockley Surface States}). Indeed, unlike
the Dirac states, they require fine tuning and are present only in a
part of parameter space (Fig. 1B). Further, since these states are typically confined to the surface on the scale of a few lattice constants, they are sensitive to surface disorder potential and, unlike
Dirac surface states, are easily localized by the disorder.

Dirac surface states arise in diverse fields, from high-energy
to solid-states physics. Early work on Dirac surface states in a
periodic potential dates back to the 1960s
\cite{davison1969relativistic,subramanian1972relativistic}. These studies have led to interesting developments in nuclear and particle
physics such as the MIT bag model and neutrino billiards \cite{bogolioubov1968modele,chodos1974baryon, 
berry1987neutrino,jaffe1989bound}. 
Recently, the interest in this problem has been renewed with
the advent of graphene and other Dirac materials \cite{mccann2004symmetry, akhmerov2008boundary}.
However, while a number of important aspects of these states have
been explored for atomically clean boundaries 
\cite{yao2009edge,latyshev2014transport,leykam2016edge,volkov2016surface}, 
the ease
with which Dirac surface states emerge, as well as their ubiquitous character, has remained unnoticed. Below we discuss the
mechanism underlying this behavior and address the key proper-
ties such as robustness, stability, and immunity to disorder. Our
work complements recent studies of topological semimetals \cite{kharitonov2017universality}.

\section{Surface states: general theory}

We first consider the general properties of Dirac surface states in
a 3D solid and then focus on the case of a graphene monolayer.
We analyze a Dirac Hamiltonian in 3D with boundary conditions
of a general form
\be\label{eq:H_Dirac_general}
H = \boldsymbol{\alpha} \vec{p}v+\beta\Delta,\qquad M\psi_\alpha\bigl|_{\rm B}=\psi_\alpha\bigl|_{\rm B},
\ee
with $\boldsymbol p=-i\mathbf (\partial_x,\partial_y,\partial_z)$ the momentum operator. Here $\{\psi_\alpha\}$ is a four-component wavefunction, 
$\{\alpha_i\}$ and $\beta$ are $4\times 4$ Dirac matrices satisfying the canonical algebra $\alpha_i\alpha_{i'}+\alpha_{i'}\alpha_i=2\delta_{ii'}$, $\alpha_i\beta+\beta\alpha_i=0$, $\alpha_i^2=\beta^2=1$. 
The parameters $v$ and $\Delta$ in \eqref{eq:H_Dirac_general} describe the 3D Bloch band structure near the Dirac band crossing. The matrix $M$ is a unitary Hermitian operator constrained
by time-reversal symmetry and current conservation \cite{mccann2004symmetry,akhmerov2008boundary}:
\be \label{eq:M_properties}
[\mathcal T,M]=0, \qquad I_{\rm B}M+M I_{\rm B}=0,
\ee
where  $\mathcal T$ is the time reversal operator, and $I_{\rm B}$ is the  current component  normal to the boundary. 

The form of these boundary conditions and the constraints on $M$ in Eq.$\,$\eqref{eq:M_properties} can be understood as follows. First, since the Dirac
equation is first order in derivatives, the boundary condition must
be stated in terms of $\psi$ alone without invoking derivatives of $\psi$. The most general
boundary condition can therefore be written as $(M-\hat 1)\psi\bigl|_{\rm B}=0$ with $M$ a suitably chosen $4\times 4$ matrix with two eigenvalues equal to $+1$. Every eigenvalue equal $+1$ yields a scalar relation between the  components of $\psi$, providing a convenient encoding of the boundary conditions in a matrix form. A considerable simplification can be achieved, without any loss of generality, by choosing $M$ to be a Hermitian
matrix with eigenvalues $+1$, $+1$, $-1$ and $-1$ (the eigenvectors corresponding to $-1$ eigenvalues do not impact the boundary conditions in any way). The form of matrix $M$ is further constrained by the requirements due to time-reversal symmetry and
probability current conservation (unitarity of scattering at the
boundary requires that the eigenvectors of $M$ with $+1$ eigenvalues give current which is  tangential to the boundary). These constraints are expressed by the first and second relation in Eq.$\,$\eqref{eq:M_properties}, respectively (for a more detailed discussion see Refs. \cite{mccann2004symmetry,akhmerov2008boundary}).

The task of finding surface states from the Dirac Hamiltonian of a general form, Eq.$\,$\eqref{eq:H_Dirac_general}, can be simplified by transforming it to a 1D Dirac problem as follows. Without loss of generality, we take the system boundary to be a 2D plane perpendicular to the $x$ direction. Using translation invariance along $y$ and $z$, we use Fourier transform, seeking the states of the form $\psi(x) e^{ik_yy+ik_zz}$. 
Assuming the system to be homogeneous
and isotropic in the $y$-$z$ plane, we can choose a new coordinate system such that $\vec k\parallel \hat y$. 
This amounts to a unitary transformation
of the spinor wavefunction and Dirac matrices, $\alpha_i'=U^{-1}\alpha_i U$, $\beta'=U^{-1}\beta U$, such that
\be
\begin{matrix}
\alpha'_1 = \alpha_1,\\
\beta' = \beta,
\end{matrix} \qquad
\begin{matrix}
\alpha'_2 = (\alpha_2k_y+\alpha_3k_z)/k,\\
\alpha'_3 =  (\alpha_2k_y-\alpha_3k_z)/k,
\end{matrix}
\ee
where $k=\sqrt{k_y^2+k_z^2}$. 

To simplify the analysis, we use, without loss of generality, an
asymmetric representation for the transformed matrices
\be\label{eq:matrices_asymmetric}
\alpha'_{1,2}=\left(\begin{matrix} \sigma_{1,2} & 0 \\ 0 & \sigma_{1,2} \end{matrix}\right)
,\ \
\alpha'_3 =\left(\begin{matrix} 0 & \sigma_3 \\ \sigma_3 & 0 \end{matrix}\right)
,\ \ 
\beta'=\left(\begin{matrix} \sigma_3 & 0 \\ 0 & -\sigma_3 \end{matrix}\right)
,
\ee 
writing it in a shorthand notation as 
$\alpha'_1 = \tau_0\s_1$, $\alpha'_2 = \tau_0\s_2$, $\alpha'_3 = \tau_1\s_3$, $\beta' = \tau_3\s_3$, where $\tau_i$ and $\sigma_i$ are $2\times2$ Pauli matrices
and $\tau_0$ is a unit $2\times2$ matrix (from now on, we suppress it for brevity). This transforms the 3D Dirac equation into a quasi-1D problem $H\psi_k (x)= \ve_k\psi_k(x)$ on a halfline $x\geq0$:
\be\label{eq:2d_ham}
H = -iv\partial_x\s_1+vk\sigma_2+\Delta\tau_3\sigma_3
, \quad 
M\psi_k(0)=\psi_k(0)
.
\ee
Surface states in 3D correspond to the one-dimensional bound states obtained from the Hamiltonian in Eq.$\,$\eqref{eq:2d_ham} (see Fig. 2C).

The advantage of this representation, in particular the choice of $\alpha_i'$, is that it allows to bring the matrix $M$ to a tractable form. All possible situations that may occur near the surface are
parameterized by different choices of the matrix $M$, whereas the Hamiltonian $H$ takes a standardized form. 
This provides a vehicle for classifying different types of behavior near the surface, parameterized by the $M$-manifold. The block representation in Eq.$\,$\eqref{eq:matrices_asymmetric} greatly facilitates this analysis. 
In this representation  the operators in Eq.$\,$\eqref{eq:M_properties} take the form $I_{\rm B} = v\alpha'_1=v\sigma_1$ and $\mathcal T=\tau_2\sigma_2 \mathcal K$, where $\mathcal K$ is complex conjugation. 
The  constraints on $M$ given in Eq.$\,$\eqref{eq:M_properties} can now be resolved as follows \cite{mccann2004symmetry,akhmerov2008boundary}. The relation $I_{\rm B}M+M I_{\rm B}=0$ implies that $M\sim \vec n\cdot\sigma$ where $\vec n$ s a vector tangential to the boundary or
a linear combination of several such terms. Combining it with the
first relation in Eq.$\,$\eqref{eq:M_properties} gives 
\be
M = (\nu\cdot\tau)(n\cdot\sigma),\qquad n_x=0.
\ee
where $n$ and $\nu$ are three-component unit vectors.  
The Hamiltonian in Eq.$\,$\eqref{eq:2d_ham} is invariant under unitary transformations of valley
matrices $\tau_i$ preserving $\tau_3$. Therefore, $M$ can be fixed, without loss of generality,
by specifying only two real phases $\theta_n$ and $\theta_\tau$:
\be \label{eq:general_Bcond}
M(\theta_\tau,\theta_n) = (\tau_3\cos\theta_\tau+\tau_2\sin\theta_\tau)(\sigma_3\cos\theta_n+\sigma_2\sin\theta_n).
\ee
The 1D problem in Eq.$\,$\eqref{eq:2d_ham} can now be solved for $M$ of a general form detailed in  Eq.$\,$\eqref{eq:general_Bcond}, giving states that decay into the bulk as evanescent waves $\psi^s_k\sim\exp(-\mu_{k,s}x)$ ( Fig.2C). The energies $\ve_{k,s}$ and 
the decay parameters $\mu_{k,s}$ obey 
\be
\label{eq:spectrum_DS} 
\begin{array}{l}\epsilon_{k,s} = \Delta\cos\theta_\tau\cos\theta_n+sK\sin\theta_n \\
\mu_{k,s} = \Delta\cos\theta_\tau\sin\theta_n-sK\cos\theta_n
\end{array}
,\quad
s=\pm 1
,
\ee
where $K = (v^2k^2+\Delta^2\sin^2\theta_\tau)^{1/2}$ and $s$ labels two possible dispersion branches (see \textit{Supplementary Information: Dirac Mode Dispersion} and Fig. 1). 
Solutions confined to the surface exist only when $\mu_{k,s}>0$.
The
resulting modes and their evolution upon changing boundary
conditions are illustrated in Fig. 2 A and B (see discussion below).

The dependence of the dispersion in Eq.$\,$\eqref{eq:spectrum_DS} on the angles $\theta_n$, $\theta_\tau$ parameterizing $M$ indicates that the surface modes exist for generic $\theta_n$, $\theta_\tau$ values, disappearing only for a subset of measure zero.  
Possible dispersion types, comprising either two branches or a single branch, are
shown in Fig. 1. The modes lie inside and outside the bandgap of the bulk spectrum. The two branches, when present, are separated by a minigap which closes at particular values $\theta_n$, $\theta_\tau$. 
\begin{figure}[t] 
\label{fig:spectrum}
\center{\includegraphics[width=1\linewidth]{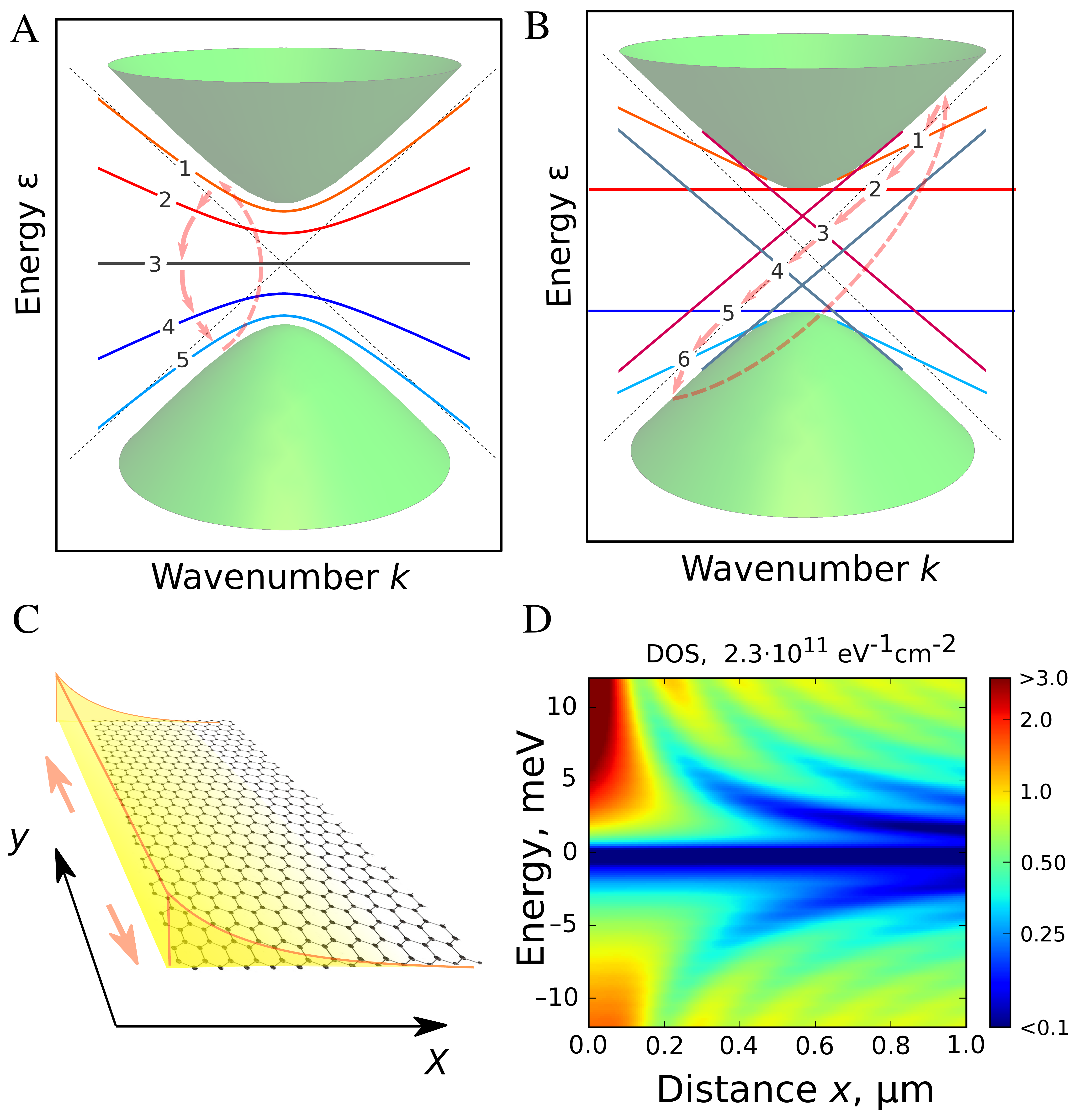}}
\caption{
(\textit{A} and \textit{B}) Surface states in monolayer graphene generated by surface potential for (\textit{A}) armchair boundary conditions and (\textit{B}) zigzag
boundary conditions. Mode dispersion changes in a cyclical manner $\dots 1 \rightarrow 2 \rightarrow 3 \rightarrow 4 \rightarrow 5 \rightarrow 6 \rightarrow 1 \dots$ with the increase of the effective potential strength $\theta_{V}$ (see \eqref{eq:phase_shift} and accompanying discussion). (\textit{C}) Edge modes are
confined at the boundary and propagate along it in both directions, as indicated by arrows. (\textit{D}) The density of states (DOS) as a function of energy and
distance from zigzag edge, with DOS in the bulk far from the edge subtracted to enhance contrast. Shown are results for case \textit{i} in Eq.$\,$\eqref{eq:zigzag_armchair_edges} for the phase-shift value $\theta_{V} = -\pi/4$. The bright peak in DOS near the edge $x = 0$ at positive energies is due to surface states. The bright peak in DOS near the edge $x=0$ at positive energies is due to surface states. The surface states contribution is embedded in a family of Friedel oscillations dispersion as $x\sim \hbar v/\epsilon$. 
}
\end{figure} 
Notably, the surface modes are present for both $\Delta \ne0$ and $\Delta =0$ i.e. 
for gapped and gapless bulk bands. 
In the latter case the modes lie outside the bulk Dirac continuum $|\epsilon|>v|k|$ and have linear dispersion of the form $\epsilon=-v\sin\theta_n |k|$. 
This gives propagation velocity of $v\sin\theta_n$. The reduction in velocity compared to the bulk velocity value provides a clear experimental signature of 
surface modes. 

\section{A relation to the Jackiw-Rebbi bound states}
To better understand the unique properties of the bulk Bloch states which enable surface states we sketch a relation between
our problem and the seminal JR problem of the midgap states
of the 1D Dirac operator with a sign-changing mass. As a first
step we perform a similarity transformation that brings $M$ to a standardized form by moving all the complexity of the problem from the boundary conditions into the transformed Hamiltonian (see \textit{Supplementary Information: Transformation to the Universal Boundary Conditions}). 
The transformation is generated by a $4\times 4$ unitary matrix that is position independent (but in general is $k$-dependent), giving two decoupled  $2\times2$ Hamiltonians
\be\label{eq:ham_w_simple_M}
H = \left(\begin{matrix}
H_+ & 0\\
0 & H_-\\
\end{matrix}\right),\qquad 
M = \left(\begin{matrix}
\sigma'_2 & 0\\
0 & \sigma'_2\\
\end{matrix}\right)
\ee
on the half-line $x\geq0$ in the new (generally, $k$-dependent) basis,
\be\label{eq:halfline_JR}
H_s = -iv\partial_x\s_1'+\ve_{k,s}\sigma_2'+\mu_{k,s}\sigma_3',\quad  s=\pm1
,
\ee
where $\ve_{k,s}$ and $\mu_{k,s}$ are defined in Eq.$\,$\eqref{eq:spectrum_DS}. 
The matrices $\sigma_i'$ are (in general, $k$- and $s$-dependent) superpositions of  $\sigma_i$. 

The $2\times2$ block structure of the transformed Hamiltonian can now be used to solve the boundary value problem. This is done most easily by extending the problem on the halfline $x>0$ to that on a full line $-\infty<x<\infty$, described by a 
Hamiltonian with a mass kink:
\be\label{eq:modified_JR}
H^{\rm R}_s = -iv\partial_x\s_1'+\ve_{k,s}\s_2'+\mu_{k,s}\sigma_3'{\rm sgn}(x) 
\ee
To identify the eigenstates which lie in the “physical” subspace of the doubled Hilbert space, we note that the Hamiltonian possesses reflection symmetry $[H^{\rm R}_s,\mathcal R]=0$, where $\mathcal R = \sigma_2 \mathcal I$ and $\mathcal I$ is 
spatial inversion $x\rightarrow -x$. 
The solutions of the problem \eqref{eq:halfline_JR} are given by the $\mathcal R$-symmetric eigenstates of $H_s^{\rm R}$ satisfying $\mathcal R \psi(x) = \psi(x)$, projected on $x> 0$.

This representation helps us to understand the robustness of surface states. It is instructive to treat $\mu_{k,s}$ as a fixed mass and use $\ve_{k,s}$ as a tuning parameter. For $\ve_{k,s}=0$ Eq.$\,$\eqref{eq:modified_JR} 
the canonical JR problem, yielding zero-mode eigenstates which at the same time are eigenstates of $\sigma_2'$ \cite{jackiw1976solitons}. Upon varying $\ve_{k,s}$ in Eq.$\,$\eqref{eq:modified_JR} 
these states remain bound to the surface albeit with
a shifted energy $\epsilon=\ve_{k,s}$.
This energy, taken as a function of $k$,
defines the dispersion of surface states.
 For $|\ve_{k,s}|\geq \mu_{k,s}$ the bound states cease to exist. 
The $k$ dependence of $\ve_{k,s}$ and $\mu_{k,s}$ is such
that the bound state may disappear in a finite range of $k$ but persist at large enough $k$ (with the exception of a measure-zero subset of $\theta_n$ and $\theta_\tau$ shown in Fig. 1A). 

\section{Edge states in graphene}
This general discussion has direct implications for graphene, the Dirac material best studied to date. In monolayer graphene, $\s_i$ and $\tau_i$ are $2\times 2$ matrices representing pseudospin and valley
degrees of freedom, respectively.
Pristine graphene is gapless with the carrier velocity 
$v\sim 10^6$ m/s. A gap as large as  $\Delta \sim 30$ meV can be created in graphene/hexagonal boron nitride superlattices. However, as discussed above, the gap has no direct significance for the existence of surface states. The types of states depend solely on the boundary conditions, i.e., the values of
the phases $\theta_n$ and $\theta_\tau$ parameterizing $M$,  which depend on the
symmetries and edge structure.

Particle-hole symmetry $\mathcal C$, if present, generates universal values $\theta_n$ and $\theta_\tau$ \cite{akhmerov2008boundary};  
namely, surface states are reduced to just two
distinct types, isomorphic to those found for crystalline zigzag and armchair edges. The allowed values are
\be\label{eq:zigzag_armchair_edges}
{\rm \it (i)}\ \theta_n=0,\pi,\ \theta_\tau =0,\pi
;\quad
{\rm \it (ii)}\ \theta_n=\pm\frac{\pi}2,\ \theta_\tau =\pm\frac{\pi}2
.
\ee 
Boundary conditions in these two cases are given by $M_1 = \pm \tau_3\sigma_3$ and $M_2  = \pm\tau_2\sigma_2$, respectively \cite{nakada1996edge}.  
In the case \textit{i} surface states form a flat band that touches one of the bulk bands bottom or top, $\ve=\pm\Delta$. In the case \textit{ii} there are no surface states. 
However, as we now
show, these restrictions are lifted for realistic non-particle-hole-symmetric edges, allowing the phases $\theta_n$ and $\theta_\tau$ to take generic nonuniversal values. 

\begin{figure}[t] 
\label{fig:schematics}
\center{\includegraphics[width=1\linewidth]{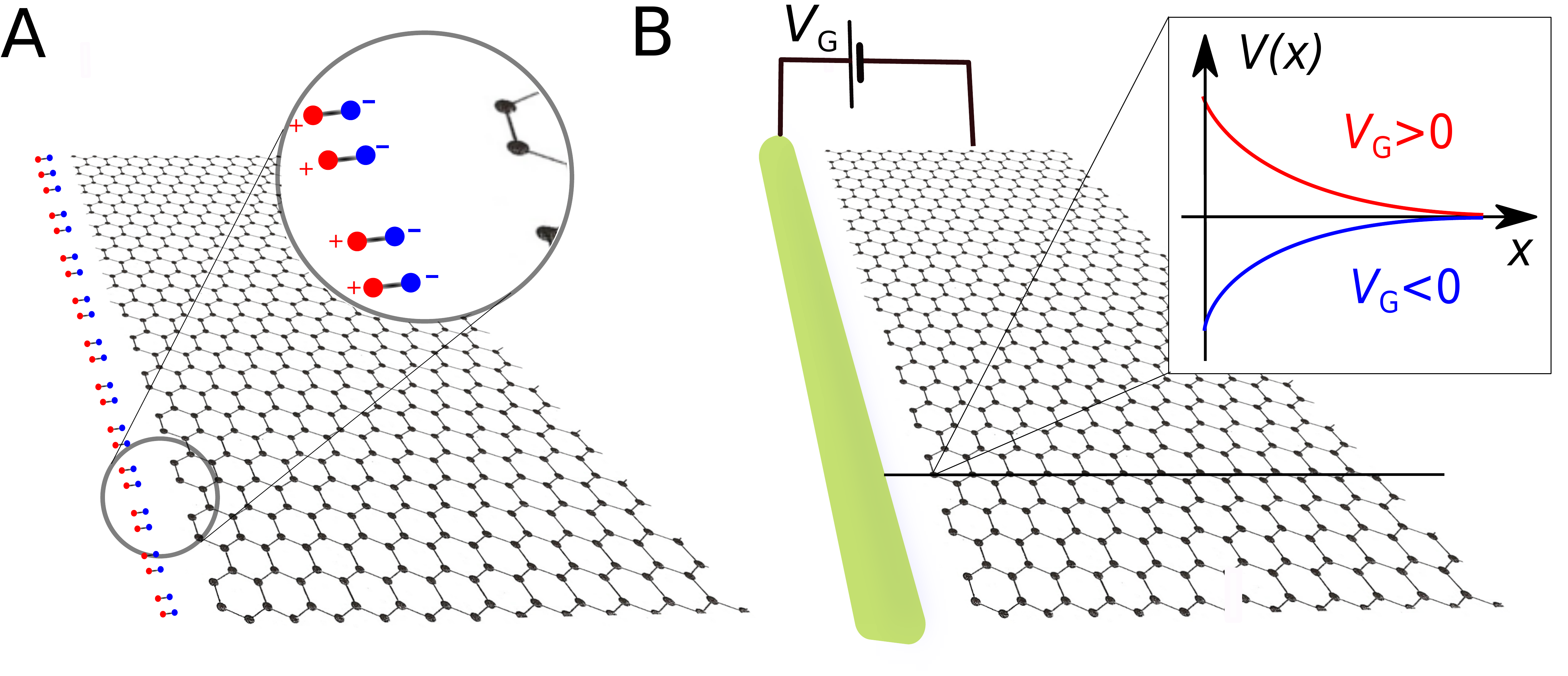}}
\caption{ 
(\textit{A}) Armchair edge in graphene monolayer. Hydrogen passivation produces atomic-scale dipoles which create electrostatic potential at the edge. (\textit{B}) Potential at the edge can be tuned by a side gate. Positive or negative potential attracts to the edge electrons or holes, respectively, modifying the dispersion of edge states as shown in Fig. 2 (Eq.$\,$\eqref{eq:phase_shift} and accompanying discussion).
}
\end{figure} 

The  $\mathcal C$ symmetry can be lifted by an edge potential that creates Dirac band bending near the edge.
 The edge potential can either occur naturally due to 
e.g. edge reconstruction \cite{koskinen2009evidence} or hydrogen passivation \cite{kobayashi2006edge,park2011formation}  or induced externally by a side gate, as illustrated in Fig.3 A and B.  
Focusing on the first case,
we consider electrostatic potential localized near the edge at a lengthscale of a few atomic spacings $r_0\sim 1$ nm:
\be\label{eq:H_with_potential}
H(x) = H+V(x), \qquad \lim_{x\rightarrow \infty}V(x)=0
\ee
Potential $V(x)$ affects states 
only in the vicinity of the edge. It is therefore convenient 
to incorporate the effect of $V(x)$ into the boundary conditions. 
This can be achieved by introducing a transfer matrix $T(x_1,x_2)$ connecting the wavefunction values, separately at each $k$, at adjacent points $x_1$ and $x_2$:
\be
\psi_k(x_1) = T(x_1,x_2)\psi_k(x_2),\qquad 0<x_2<x_1<\infty,
\ee
where $\psi_k(x)$ is obtained from the 
Dirac equation with the Hamiltonian in Eq.$\,$\eqref{eq:H_with_potential}.  
The 
transfer matrix can be obtained by integrating the Dirac equation over $x$, 
\be\label{eq:SM_transferT_V}
T(x_1,x_2)={\rm Xexp}\int_{x_2}^{x_1}dx\frac{i}{v}\Bigl(D_k-\sigma_1  V(x) \Bigl),
\ee
where $D_k = i\sigma_1\ve+\sigma_3vk-\tau_3\sigma_2\Delta$ and
${\rm Xexp}$ denotes an $x$-ordered exponential. 
Assuming that the term $V(x)$ gives negligible contribution for $x>r_0$, we can approximate the transfer matrix as a product of a free-particle contribution and a boundary term,
\be\label{eq:divide_operator}
T(x,0) = T(x,r_0)T(r_0,0)\approx T_0(x,r_0)T(r_0,0)
\ee
where $T_0$ is a transfer matrix for zero potential $ V(x)=0$. The boundary contribution $T(r_0,0)$ can be expressed in a closed form through $V(x)$ when the potential radius $r_0$ is much smaller than electron wavelength $\lambda=\hbar v/\epsilon$. This is achieved by writing $\psi_k(x>r_0) = T_0(x,r_0)\psi_k(r_0)$, shifting the boundary position to $x = r_0$, and writing the boundary conditions as
\be\label{eq:new_Bc}
\psi_k(r_0)= M_V\psi_k(r_0)
,\quad 
M_V = \Theta M\Theta^{-1},
\ee
where we denote $\Theta = T(r_0,0)$. 
Under the conditions $kr_0\ll 1$ and $\Delta r_0/\hbar v\ll 1$ we can ignore the first term inside ${\rm Xexp}$ in Eq.$\,$\eqref{eq:SM_transferT_V}. 
Approximating $\int_0^{r_0}V(x)dx\approx \int_0^{\infty}V(x)dx$ then gives a matrix $M$ that describes the boundary condition altered by $V(x)$:
\be\label{eq:phase_shift}
M_V = M(\theta_\tau,\theta_n+\theta_{V}), \qquad \theta_{V}=\frac2v\int_0^{\infty}V(x)dx.
\ee
This simple result is valid as long as the edge potential width $r_0$
is small compared with the electron wavelength in the bulk.

We note parenthetically that the latter condition restricts the validity of our approach to short-range edge potentials and narrow-gap Dirac band structures such that $r_0\ll \lambda=\hbar v/\Delta$. In
wide-gap band structures the interaction of carriers with the crystal surface is in general not described by a simple scalar potential model. In addition, long-range potentials can produce many bound states at the edge and thus create many surface modes.

To understand the impact of the edge potential on the edge states dispersion we consider the setup of Fig.3B wherein $V(x)$ is tuned by a side gate. 
Through varying $\theta_{V}$ the edge states dispersion changes in a complex way, 
as illustrated in Fig. 2 A and B for the armchair and zigzag edge. 
The armchair edge hosts
a one-branch mode with relativistic dispersion
\be\label{eq:armchair_tunable}
\ve_k = -\sqrt{k^2v^2+\Delta^2}\cos\theta_{V}. 
\ee
Interestingly, the mode in Eq.$\,$\eqref{eq:armchair_tunable}, despite its relativistic appearance, has no $\mathcal C$-symmetric counterpart; i.e., it does not obey particle-hole symmetry. Furthermore, the dispersion acquires a flat-band character at $\theta_{V}=\pm\pi/2$. 

The solution for the zigzag edge features a more complex
behavior. For each $\theta_V$ value, the edge modes contain two distinct branches, propagating to the right and to the left, as illustrated in Fig. 2B.
Modes occurring at $0<\theta_{V}<\pi$ for zigzag edges with 
$\theta_n=0$ and at $-\pi<\theta_{V}<0$ for 
zigzag edges with $\theta_n=\pi$ 
(cases 3 and 4 in Fig. 2B) span both positive and negative energies. Upon
variation of $\theta_{V}$ they sweep through the bulk bandgap. For other
$\theta_{V}$ values, the modes also consist of two branches; however, their
energies are either above or below the bulk gap (cases 1 and 6 in Fig. 2B). The dispersion becomes flat at $\theta_{V}=0,\pi$ (cases 2 and 5 in Fig. 2B). Since according to Ref.$\,$\cite{akhmerov2008boundary} the zigzag boundary condition with $\theta_{V}=0$ describes a generic $\mathcal C$-symmetric lattice termination in monolayer graphene, our solution for $\theta_{V}\ne 0$ describes modes for a generic gated graphene edge. 

The contribution of surface states to spatially resolved DOS is illustrated in Fig. 2D (for derivation see \textit{Supplementary Information: Spatially Resolved Density of States}). Surface states give rise to an
enhanced DOS near the boundary for one type of carriers,
electrons or holes, depending on the $\theta_V$ value. In Fig. 2D the contribution of surface states is seen as a high-DOS region at positive energies, embedded into the family of Friedel oscillations dispersing as $r\sim \hbar v/\epsilon$.

\section{The role of disorder}
An interesting aspect of Dirac surface states is their weak interaction with surface disorder. Realistic crystal boundaries often feature strong disorder potential, arising due to dangling bonds and other defects, which impedes transport along the surface. Suppression of conduction by surface disorder is typically quite strong for the Tamm–Shockley states. However, Dirac surface states are to a great extent protected from surface scattering due to their small overlap with the surface disorder. This behavior is reminiscent of the carrier dynamics in GaAs/AlAs
quantum wells where the mobility increases drastically with the well width due to a rapid mean free path growth $\ell \sim w^n$, with large $n$ 
\cite{gold1986metal}.  
Since the well width is much greater than the interface roughness scale, carriers can diffract around the interface
disorder. In our case, a similar diffraction-enhanced conduction occurs since the width of the surface states, which defines
their extent into the bulk and is on the order of the bulk
wavelength $\sim\lambda_F$, is much larger than atomic surface roughness. Effectively, in this case the mode width takes on a role analogous to the width of quantum wells. This allows Dirac surface states to propagate quasi-ballistically with negligible surface
scattering. 

To illustrate the effect of backcattering suppression by electron-wave diffraction around surface disorder, we consider gaussian short-range correlated disorder at the graphene edge,
\be \label{eq:random_potential}
H_{\rm dis} = H+\xi(y)\delta(x)
,\quad 
\la \xi(y)\xi(y')\ra_{\rm dis}=\alpha\delta(y-y')
\ee 
with $\alpha$ the disorder strength parameter. While the disorder spectrum is broad band, only the harmonics comparable to carrier wavelengths scatter efficiently \cite{unuma2003intersubband} whereas the contribution of other harmonics 
is relatively weak \cite{kamburov2016interplay}. 

In the limit of a weak disorder the mean free path can be evaluated by perturbation theory (see \textit{Supplementary Information: Disorder at the Edge}). Here we discuss the results for the zigzag surface state (case \textit{i} in Eq.$\,$\eqref{eq:zigzag_armchair_edges}). In this case, the mean free path is
\be\label{eq:mean_free_path}
\ell = \frac{\lambda_{\rm B}^2}{\zeta},\qquad \zeta = \frac{8\pi^2\alpha}{\hbar^2 v^2}\cot^2\theta_{V}
\ee
where $\lambda_{\rm B}=2\pi\hbar v\sin\theta_{V}/\ve_F$ is the carrier wavelength in the edge mode, $\ve_F $ is Fermi energy.  
The lengthscale $\zeta$, proportional to disorder strength, 
can be estimated as $\zeta\sim \alpha/\hbar^2v^2\sim U_0^2 a^3/\hbar^2v^2 \sim 1 \text{ nm}$,
where $U_0\sim 1$ eV is an atomic-scale potential and 
$a\sim1$ nm is surface roughness. For $\zeta\ll\lambda_{\rm B}$ Eq.$\,$\eqref{eq:mean_free_path} predicts mean free path values much greater than the carrier wavelength; the dimensionless parameter $\lambda_{\rm B}/\zeta$ describes the effect of scattering suppression by diffraction. As an illustration, for $ \theta_{V}\approx 1$ and wavelength of order $\lambda_{\rm B} \sim 100$ nm 
we obtain the value $\lambda_{\rm B}/\zeta\approx 100$, giving the diffraction-enhanced mean free path 
as large as
$L\sim 10^{2}\lambda_{\rm B}= 10^4\,{\rm nm}$. 

We note that localization effects may become important if the disorder is strong enough. In our case, since disorder is mainly at the surface, the behavior is expected to be quite different for electron energies inside and outside the bulk energy gap. In the first case, electron states with energies within the bulk gap reside near the surface.
These states couple to surface disorder relatively
strongly and may become localized.
In
the second case the states at the surface will hybridize with the states in the bulk, which suppresses localization due to surface
disorder. In addition, as discussed above, the slow decay of electron states from the surface into the bulk gives the surface states a large width that allows electrons to diffract around surface disorder. Such diffraction also suppresses localization. For quasi-1D surface states, such as those in graphene, the 1D mean free path provides a good estimate for localization length at the energies in
the bulk gap. For 2D surface states, on the other hand, the localization length is expected to be much longer than the mean free path estimated perturbatively. The latter in this case sets only a lower bound for localization length.

\section{The effect of magnetic field}
Experimental detection of surface states by conventional transport techniques can be challenging since the signatures of surface
states are often obscured by the continuum of bulk states (the overlap of bulk and surface states contributions to the density of states is illustrated in Fig. 2D). Here we consider a different approach relying on the Landau-level spectroscopy in a magnetic field applied perpendicular to the surface. The signatures of Landau levels of the states in a 3D bulk are usually softened by the momentum dispersion in the direction along the field. In contrast, the spectrum of the 2D surface states will be discrete. Therefore, while both the bulk states and the surface
states produce Landau levels, the spectral features such as, e.g., the tunneling density of states measured by scanning tunneling microscopy will be dominated by the surface states. 

To study the effect of magnetic field, we use a simple model of electrons confined by a 2D delta-function sheet potential of the strength proportional to $\theta_{V}$ (Eq.$\,$\eqref{eq:phase_shift}):
\begin{equation}
H = \sigma_1 vp_x+\sigma_2 v\left(p_y+eBz\right)+\tau_2\sigma_3vp_z+\tau_3\sigma_3\Delta +v\theta_{V}\delta(x)
\end{equation}
The states confined near the $x=0$ plane can be found as evanescent solutions for $x>0$ and $x<0$,  $\psi\sim\exp(ip_xx+ip_yy-\mu|x|)$ (see \textit{Supplementary
Information: Surface States in Magnetic Field}). For $B = 0$, the spectrum of this model coincides with the spectrum of surface states for armchair boundary conditions:
\be\label{eq:free_delta_pot}
\ve_k = -{\rm sgn}(\theta_{V})\cos\theta_{V}\sqrt{v^2k^2+\Delta^2},\qquad k^2 = p_y^2+p_z^2
\ee
In a nonzero magnetic field $B$ , we obtain discrete nondispersing levels resembling Landau levels of 2D Dirac particles:
\be\label{eq:discrete_delta_pot}
\ve_n = -{\rm sgn}(\sin\theta_{V})\cos\theta_{V} \sqrt{2veBn+\Delta^2},\qquad n=0,1,2\dots
\ee
Interestingly, similar to the $B=0$ solution in Eq.$\,$\eqref{eq:free_delta_pot}, the discrete levels exist only for one sign of energy, positive or negative.

The discrete character of the surface Landau levels as well as their striking lack of particle–hole symmetry provides a direct and simple diagnostic of the surface states. Further evidence can be obtained using the property of surface states to be tunable through changing the surface potential by side gates (Fig. 3 and Eq.$\,$\eqref{eq:phase_shift}). Due to a periodic dependence on the potential strength,
the electron-hole asymmetry can be inverted by reversing the potential sign or by applying a stronger potential.

In conclusion, our key finding is that surface states are a natural attribute of a Dirac band structure, appearing in a robust manner for generic boundary conditions. The surface states feature a number of interesting and potentially useful properties. In particular, we predict that these states are insensitive to surface imperfections: By diffracting around surface disorder electron waves can propagate ballistically with abnormally long mean free path values. These states can coexist with the bulk
states or appear within the bulk bandgap; their dispersion can be tuned by gate potential or by magnetic field, giving rise to unique signatures amenable to a variety of experimental probes. 

\textbf{Acknowledgements}. We acknowledge support of the Center for Integrated Quantum Materials under NSF Award DMR-1231319; the MIT Center for Excitonics; the Energy Frontier Research Center funded by the US Department of Energy, Office of Science, Basic Energy Sciences under Award de-sc0001088; and Army Research Office Grant W911NF-18-1-0116.

\bibliography{bibliography}
\clearpage

\pagebreak

\setcounter{page}{1}
\setcounter{equation}{0}
\setcounter{figure}{0}
\renewcommand{\theequation}{S.\arabic{equation}}
\renewcommand{\thefigure}{S\arabic{figure}}
\renewcommand*{\thepage}{S\arabic{page}}

\onecolumngrid

\begin{center}
{\huge$\mathbf{Supplementary \, Information}$}\\
\vspace{0.5cm}
\end{center}
\vspace{0.5cm}

\twocolumngrid

\section*{
The Tamm-Shockley surface states}

The Tamm-Shockley (TS) states are electronic states
confined near 
crystal boundaries of semiconductors [1].
These states 
can occur only for crystal boundaries satisfying specific conditions. When the TS states exist they are typically localized within an atomically thin layer near the surface. We illustrate this behavior with a simple model: the nonrelativistic 
Schr\"{o}dinger equation with a periodic potential that depends on the coordinate perpendicular to the surface:
\be\label{eq:quadratic_S_eq}
\nabla^2 \psi +
2m\Bigl(\ve-U(x)\Bigl)\psi=0,
\ee
where $\nabla = (\partial_x,\partial_y,\partial_z)$ is the gradient operator, $\ve$ 
and $m$ are particle energy and effective mass. For simplicity, we take 
the crystal lattice potential to be constant along the surface. 
We consider a semi-infinite crystal with the boundary placed at $x=0$, and approximate the potential $U(x)$ dependence on the coordinate normal to the boundary 
by a periodic Dirac comb potential inside the solid, at $x>0$, and a constant work function potential outside, at $x\leq 0$:
\begin{align}\label{eq:potential}
&U(x) = 
\begin{cases}
\sum_n {V}\delta(x-an)\quad &x>0 \\
U_0 \quad &x\leq0,
\end{cases}
\quad n=1,2,3...
\end{align}
where 
$a$ is a lattice spacing, ${V}>0$ is the delta function amplitude, and $U_0>0$ is the work function for the material [2]. 

Below, without loss of generality, we consider states with zero momentum component along the surface, $k_y=k_z=0$. The dependence on $k_y$ and $k_z$ can be absorbed in the energy by a shift $\epsilon\to\epsilon-(k_y^2+k_z^2)/2m$. 

The solution for Eq.$\,$\eqref{eq:quadratic_S_eq} with the potential given in Eq.$\,$\eqref{eq:potential} 
can be obtained separately for positive and negative $x$, and then combined together, accounting for wavefunction continuity at $x=0$. In the region $x<0$ the solution with energy $\ve<U_0$ is
\be
\psi_1(x) = A_1\exp(\kappa x), \quad \kappa = \sqrt{2m(U_0-\ve)},
\ee  
where $A_1$ is a complex parameter.

In the region $x>0$ the system is described by Bloch wavefunction $\psi_2(x)$ with a Bloch wavenumber $k$, satisfying $\psi_2(x+a)  = e^{ika} \psi_2(x)$. For the Dirac comb potential, Bloch theorem leads to a condition connecting the wavenumber and the energy (see Refs. [3,4] for details):
\be\label{eq:dc_condition}
\cos k a = p\frac {\sin \xi a}{\xi a}+\cos\xi a\\
\ee
where $p = am{V}$, and $\xi = \sqrt{2m\ve}$. When the right-hand side of Eq.$\,$\eqref{eq:dc_condition} is smaller or equal to one, $\ve$ represents energy of a plane wave with wavenumber $k$. 
When the right-hand side of Eq.$\,$\eqref{eq:dc_condition} is greater than one, solutions correspond to complex $k = i\mu+n\pi$, with $\mu$ real and $n$ integer. For infinite crystal, the energies with complex $k$ give no physical solutions and lie in a band gap. However, in the case of a semi-infinite crystal $x>0$, solutions with $\mu>0$ describe confined states at the boundary. Let us focus on the latter case and consider the solution in the first cell $0<x<a$ [3,4]:
\be
\psi_2(x) = A_2(e^{i\xi x}+\beta e^{-i\xi x}),\quad \beta = -\frac{1-e^{-i(k-\xi)a}}{1-e^{-i(k+\xi)a}}
\ee
where $A_2$ is a complex parameter. Here the expression for $\beta$ is found by using the Bloch condition and matching the wavefunction values in the first and second unit cells.

The continuity condition at the boundary $x=0$ leads to a pair of equations for the  parameters $A_1$ and $A_2$:
\be
\begin{split}
\psi_1(0) = \psi_2(0):\quad  &A_1 = A_2(1+\beta)\\
\frac d{dx}\psi_1\Bigl|_{x=0} = \frac d{dx}\psi_2\Bigl|_{x=0}:\quad  &A_1\kappa = iA_2\xi (1-\beta)
\end{split}
\ee
These linear equations have a non-zero solution 
when the corresponding determinant vanishes, giving 
\be\label{eq:dc_continuity}
e^{-\mu a+i\pi n} = \sqrt{q^2-\xi^2}\frac {\sin \xi a}{\xi a}+\cos\xi a
\ee
where $q = \sqrt{2mU_0}$.

Another constraint for the parameters $\mu$ and $\xi$ is obtained by plugging $k=i\mu+n\pi$ in Eq.$\,$\eqref{eq:dc_condition}, which gives
\be\label{eq:dc_complex_solution}
(-1)^n\cosh \mu a = p\frac {\sin \xi a}{\xi a}+\cos\xi a
. 
\ee
Subtracting \eqref{eq:dc_continuity} from this expression gives a relation
\be\label{eq:dc_mu_condition}
(-1)^n\sinh \mu a = (p-\sqrt{q^2-\xi^2})\frac {\sin \xi a}{\xi a}
.
\ee
Comparing to Eq.$\,$\eqref{eq:dc_complex_solution} we see that,  
since $\cosh (\mu a) \geq 1$, the sign of ${\sin (\xi a)}/{\xi a}$ for any solution coincides with $(-1)^n$. 
Therefore, a solution confined to the boundary, described by $\mu>0$, exists only when  $p>\sqrt{q^2-\xi^2}$ in Eq.$\,$\eqref{eq:dc_mu_condition}. This condition can be satisfied provided that $p>q$. In terms of the parameters of the Hamiltonian, the existence condition is written as
\be\label{eq:LP_condition}
{V}>\sqrt{\frac2m U_0}.
\ee
The phase diagram obtained from Eq.$\,$\eqref{eq:LP_condition} is shown in Fig.1B of the main text. In this figure the following notations are used: $t=V/a$, $J_0=2/ma^2$. If the condition in Eq.$\,$\eqref{eq:LP_condition} is not fulfilled, the TS states do not exist.

This example illustrates that for carriers in a single Bloch band the existence of surface states requires fine tuning of the crystal lattice parameters. The situation is quite different for a Dirac band (i.e. two Bloch bands with a narrowly avoided crossing). In this case, as discussed in the main text, surface states appear in a generic manner.

\section*{Dirac mode dispersion} 
 
The system described by the Hamiltonian in Eq. (5) and constrained by the boundary conditions, Eq. (6), has both bulk plane wave solutions as well as solutions confined to the surface. In this section we derive the spectrum of confined states and show that these states exist for generic boundary conditions with the exception of a set of measure zero. As an ansatz, we look for solutions in the form of evanescent waves in the region $x>0$:
\be \label{eq:SM_ansatz}
\psi_k(x) =\phi_k\exp(-\mu_k x),\qquad \mu_k = \frac1v\sqrt{\Delta^2+v^2k^2-\ve_k^2},
\ee 
where $\phi_k$ is a normalized 4-spinor that obeys the 
Dirac equation
\be\label{eq:S_equation_spinor}
\Bigl(iv\mu_k\sigma_1+vk\sigma_2+\Delta\tau_3\sigma_3\Bigl)\phi_k=\ve_k\phi_k.
\ee
The general solution of Eq.$\,$\eqref{eq:S_equation_spinor} is a superposition of $\tau_3$ eigenvectors $\phi_k = \alpha \phi_1+\beta\phi_2$,
\begin{multline}\label{eq:SM_K_Kp}
\phi_1 = 
\left(\begin{matrix}
iv(\mu_k-k),\\
\ve_k-\Delta\\
0\\
0
\end{matrix}
\right),
\qquad
\phi_2 = 
\left(\begin{matrix}
0\\
0\\
iv(\mu_k-k),\\
\ve_k + \Delta\\
\end{matrix}
\right),
\end{multline}
where $\alpha$ and $\beta$ are complex parameters to be determined below. For monolayer graphene, $\phi_1$ and $\phi_2$ correspond to $K$ and $K'$ valley polarizations.

As noted in the main text, the most general form of the spinor $\phi_k$ satisfying the boundary conditions Eq. (6) also must lie to the subspace spanned by the $M$ eigenvectors with the eigenvalue $+1$: 
\be\label{eq:SM_pp_mm}
\phi_k = A|\theta_\tau +\rangle|\theta_n+\rangle+B|\theta_\tau -\rangle|\theta_n-\rangle,
\ee
where
\be
|\theta+\rangle = 
\frac1{\sqrt{2}}\left(\begin{matrix}
\cos\theta/2\\
i\sin\theta/2
\end{matrix}\right),
\quad
|\theta-\rangle = 
\frac1{\sqrt{2}}\left(\begin{matrix}
-\sin\theta/2\\
 i\cos\theta/2
 \end{matrix}\right),
\ee
and $A$, $B$ are some complex parameters. The double-ket states in Eq.$\,$\eqref{eq:SM_pp_mm} is a shorthand notation for 
\be\label{doubleket_12}
|\theta_\tau +\rangle|\theta_n+\rangle=\frac 12 \lp \begin{array}{c} 
c_\tau\,c_n\\
i c_\tau\,s_n\\
i s_\tau\,c_n\\
s_\tau\,s_n
\end{array}
\rp
,\quad
|\theta_\tau -\rangle|\theta_n-\rangle=\frac 12\lp \begin{array}{c} 
s_\tau\,s_n\\
-is_\tau\,c_n\\
-ic_\tau\,s_n\\
c_\tau\,c_n
\end{array}
\rp
,
\ee
where we used notation $c_\tau=\cos\theta_\tau/2$, $s_\tau=\sin\theta_\tau/2$,  $c_n=\cos\theta_n/2$, $s_n=\sin\theta_n/2$. 

Together Eq.$\,$\eqref{eq:SM_K_Kp} and Eq.$\,$\eqref{eq:SM_pp_mm} form a set of four linear equations for four unknown parameters $\alpha,\beta,A,B$:
\be\label{eq:b_cond_eq_set}
\alpha \phi_1+\beta\phi_2 = A|\theta_\tau +\rangle|\theta_n+\rangle+B|\theta_\tau -\rangle|\theta_n-\rangle.
\ee
The set of equations in Eq.$\,$\eqref{eq:b_cond_eq_set} has a non-zero solution provided the corresponding $4\times 4$ matrix $R$, comprised of the spinors in Eq.$\,$\eqref{eq:SM_K_Kp} and Eq.$\,$\eqref{doubleket_12},  is degenerate. This is the case when the determinant of this matrix vanishes:
\be\label{eq: determinant}
{\rm det} R=i(\mu_k-k)(v\mu_k\sin\theta_n+\ve_k\cos\theta_n-\Delta\cos\theta_\tau)=0.
\ee
The first term in Eq.$\,$\eqref{eq: determinant} vanishes only when $\ve_k=\pm\Delta$, as seen from Eq.$\,$\eqref{eq:SM_ansatz}. Setting the second term to zero and solving the resulting equation together with Eq.$\,$\eqref{eq:SM_ansatz}, we obtain two branches of surface states. The corresponding energies $\ve_k$ and the parameters $\mu_k$ are given by 
\begin{align}\label{eq:SM_general_dispersion}
&\ve_k = \Delta\cos\theta_\tau\cos\theta_n\pm\sin\theta_n\sqrt{v^2k^2+\Delta^2\sin^2\theta_\tau},\nonumber\\ 
&\mu_k = \Delta\cos\theta_\tau\sin\theta_n\mp\cos\theta_n\sqrt{v^2k^2+\Delta^2\sin^2\theta_\tau}.
\end{align}
This provides derivation of 
Eq. (8) in the main text.

When do the relations in Eq.$\,$\eqref{eq:SM_general_dispersion} describe confined states? The necessary and sufficient condition for confinement is $\mu_k>0$, otherwise the confined states are absent.
In the limit of large momentum $vk\gg\Delta$ 
Eq.$\,$\eqref{eq:SM_general_dispersion} becomes 
\be
\ve_k = \pm vk \sin\theta_n,\qquad \mu_k = \mp vk\cos\theta_n
\ee
In this limit it is easy to see that the condition $\mu_k>0$ is satisfied for at least one of two solutions in Eq.$\,$\eqref{eq:SM_general_dispersion} for any $\theta_n\neq\pm\pi/2$. 

The confined states disappear when $\theta_n=\pm\pi/2$, since in this case $\mu_k=0$
If consider now the case $\theta_n=s\pi/2$, $s=\pm1$ for arbitrary $k$, 
The expression in Eq.$\,$\eqref{eq:SM_general_dispersion} for arbitrary $k$ and $\theta_n=s\pi/2$, $s=\pm1$, is reduced to
\be
\ve_k = \pm s\sqrt{v^2k^2+\Delta^2\sin^2\theta_\tau}, \qquad \mu_k = s\Delta\cos\theta_\tau.
\ee
In the case $s=-1$ ($\theta_n=-\pi/2$) the confined solutions are absent for $\theta_\tau \in [0,\pi]$, whereas in the case $s=+1$ ($\theta_n=\pi/2$) they are 
absent for $\theta_\tau \in [\pi,2\pi]$. These two measure-zero subsets of the $(\theta_\tau,\theta_n)$ parameter space 
are the only regions where surface states do not exist.

The graphic summary of the results of this section is shown in the Fig.1 and Fig.2 of the main text. The phase diagram in the panel A of Fig.1 shows that the Dirac surface states are present for generic boundary conditions, namely they do not require fine tuning of system parameters. This is in contrast to the Tamm-Shockley states which, as illustrated by Fig.1 (panel B) of the main text, appear in a limited region of parameters. 
An example of the the surface states spectrum in Eq.$\,$\eqref{eq:SM_general_dispersion} for $\theta_\tau = 0$ and $\theta_\tau = \pi/2$ and varying $\theta_n$ is shown on the panels A and B of Fig. 2 of the main text.

\section*{Transformation to the universal boundary conditions}

Setting boundary conditions for the Dirac equation involves specifying 
4-spinors compatible with the properties of a given boundary. Unlike the nonrelativistic 
Schr\"{o}dinger equation, for which the wavefunction may vanish at the boundary (a hard-wall boundary condition), the Dirac 4-component spinor wavefunction never vanishes at the boundary. Instead it is described as an eigenvector of a suitably defined Hermitian $4\times 4$ matrix $M$ (see Eq.(1) in the main text and accompanying discussion). The matrix $M$ encodes the properties of the boundary through the dependence on the 
phenomenological parameters $\theta_\tau$ and $\theta_n$. However, for generic boundary conditions the matrix $M$ does not commute with the Hamiltonian and thus solving the Dirac equation near the boundary is typically not a straightforward exercise. 



Here we show that a significant simplification can be achieved by performing a unitary transformation on 
the Hamiltonian and 
the matrix $M$, such that the boundary conditions are brought to a standardized form. In particular, this transformation can be chosen to make the new boundary conditions {\it completely independent} of the parameters $\theta_\tau$ and $\theta_n$. Instead, the dependence on $\theta_\tau$ and $\theta_n$ will show up in the Hamiltonian. 
These requirements are fulfilled by a $k$-dependent unitary transformation
\be\label{eq:basis_trans}
U =
\exp(-i\theta_k\tau_2)\cdot S\cdot
\exp\biggl[-\frac i2{\theta_\tau}\tau_1+\frac i2\biggl(\frac{\pi}2-\theta_n\biggl)\sigma_1\biggl],
\ee
where  $\theta_k = \arctan(k/\Delta\sin\theta_\tau)$ and $S$ is a swap matrix
\be
S = \begin{pmatrix}
 1 &0 &0 &0 \\
 0 &1 &0 &0 \\
 0 &0 &0 &1 \\
 0 &0 &1 &0
\end{pmatrix} .
\ee
 After the transformation the new boundary conditions are expressed through matrix 
\be\label{eq:newbc}
M' = UMU^\dag = \sigma_2.
\ee
Importantly, this transformation, applied to the Hamiltonian for surface states (Eq. (5) in the main text) preserves the $\tau_3$ block structure, generating a mass term that depends on $\theta_\tau$ and $\theta_n$:
\begin{multline}\label{eq:newham}
H' = UHU^\dag =  -iv\partial_x \s_1+\Bigl(\Delta\cos\theta_\tau\cos\theta_n+\tau_3 K\sin\theta_n\Bigl)\s_2+\\
+\Bigl(\Delta\cos\theta_\tau\sin\theta_n-\tau_3 K\cos\theta_n\Bigl)\s_3 = \left(\begin{matrix}
H_+& 0\\
0&H_-
\end{matrix}\right)
\end{multline}
where $K = \sqrt{v^2k^2+\Delta^2\sin^2\theta_\tau}$ and $H_\pm$ are $2\times2$ matrices.

As discussed in the main text [Eq.(11)], 
the problem of finding surface states 
is reminiscent of the Jackiw-Rebbi (JR) problem of a 1D Dirac equation with a mass kink. In the case of surface states the kink arises after the Dirac equation in a half-space is extended to the entire space by a mirror reflection. This  interesting property not only facilitates the analysis, it also provides a simple intuitive argument for the robustness of the surface states. Indeed, in the JR problem a mass kink always creates a bound state with the energy in the Dirac bandgap. While in the original JR problem the bound state has zero energy because of the charge conjugation symmetry, for a more general Dirac Hamiltonian the bound state, arising from a suitably generalized JR analysis, in general occurs at a nonzero energy. 
In the main text the relation to the JR problem helps to derive the spectrum of surface modes. In the next section we use it to calculate the corresponding contribution to the density of states.


\section*{Spatially resolved density of states} 

An easily testable signature of surface states is an enhanced density of carriers near the surface. The wavefunctions of surface states are localized near the boundary, decaying as evanescent waves into the bulk and propagating along the boundary as plane waves (see Eq.$\,$\eqref{eq:SM_ansatz}). As we will see below, this behavior translates directly into the spatial structure of the 
density of states (DOS) $\nu(\ve,x)$ 
taken as a function of energy $\ve$ and the distance from the boundary $x$. 
We show that the surface states give the dominant contribution 
to DOS in vicinity of the surface.

We will use the standard relation between spatially resolved DOS 
and the Green's function (GF) which reads 
\be\label{eq:DOS_general}
\nu(\ve,x) = -\frac i\pi\lim_{x'\rightarrow x} \Im \Tr G(x,x').
\ee
Here $G(x,x')=\la x'|\frac1{\epsilon-H+i0}|x\ra$ where by $|x\>$ we denote eigenstates of the position operator. 
The quantity $G(x,x')$ is a matrix in the Dirac spin space, so a trace over spin variables is taken to obtain DOS.

Applying the relation in Eq.$\,$\eqref{eq:DOS_general} to the Hamiltonian and the boundary conditions derived above [Eq.$\,$\eqref{eq:newbc} and Eq.$\,$\eqref{eq:newham}] we see that the DOS 
is represented as  a sum of the contributions from the two $2\times 2$ block Hamiltonians $H_\pm$: 
\be\label{eq:dos_general_formula}
\nu(\ve,x) = -\frac i\pi\lim_{x'\rightarrow x} \Im \sum_{s=\pm1}\Tr G_s(x,x') 
,
\ee
where $G_\pm (x,x')=\la x'|\frac1{\epsilon-H_\pm+i0}|x\ra$ corresponds to the blocks $H_\pm$ in Eq.$\,$\eqref{eq:newham}. 
The space of functions in which $H$ is defined and in which the inverse $\frac1{\epsilon-H_\pm+i0}$ is considered is comprised of the functions obeying the boundary conditions given in Eq.(6) in the main text. Analyzing this space, generally speaking, is a nontrivial task, which is simplified by the trick described below.


To calculate DOS, rather than applying the boundary conditions directly, it is more convenient to use the 
relation between the system with a boundary and the Hamiltonian $H_s^{\rm R}$ extended to the full space by 
a mirror reflection 
(see discussion in the main text after Eq.(11)). For $H_s^{\rm R}$ the GF can be obtained as
\be\label{eq:system_GF}
G_s(x,x') = \<x|\mathcal P\frac1{\ve-H_s^{\rm R}}\mathcal P|x'\>,\qquad x,x'\geq0
\ee
where 
$H^R_\pm$ is the full-space Hamiltonian given in Eq.(11) in the main text, and we introduced the projection operator that eliminates unphysical solutions, 
\be
\mathcal P = \frac 12(1+\mathcal R), \qquad \mathcal P^2=\mathcal P.
\ee 
Here $\mathcal R = \sigma_2 \mathcal I$ is the mirror reflection operator given by a product of 
a 1D spatial inversion $ \mathcal I:\,x\rightarrow -x$ and spin $\pi$-rotation about the $y$-axis $\sigma_2 = i\exp(-i\pi\sigma_2/2)$. 
Here and below we suppress the prime superscripts used in Eq.(11) of the main text, replacing $\sigma'_i \rightarrow \sigma_i$ throughout the calculation. 

The full-space Hamiltonian is invariant under $\mathcal R$. Indeed, the terms of $H^R_s$ containing $\sigma_1$ and $\sigma_3$,  which anticommute with $\sigma_2$, also anticommute with the spatial inversion $ \mathcal I:\,x\rightarrow -x$:
\be
\partial_x \mathcal I = -\mathcal I \partial_x, \qquad \theta(x) \mathcal I= -\mathcal I \theta(x).
\ee 
As a result, $[H^R_s,\mathcal R]=0$. 

Taking this into account allows us to get rid of the projector operators $\mathcal P$. For this we use the property $[H^R_s,\mathcal P]=0$ and express the GF as
\begin{multline}\label{eq:mirror_trick}
G_s(x,x') =  \<x|\frac1{\ve-H_s^{\rm R}}\mathcal P^2|x'\>  =  \<x|\frac1{\ve-H_s^{\rm R}}\mathcal P|x'\>\\
=\frac 12 \Bigl(\<x|\frac1{\ve-H_s^{\rm R}}|x'\> + \<x|\frac1{\ve-H_s^{\rm R}}\mathcal R|x'\> \Bigl) \\=\frac 12 \Bigl(G_s^{\rm R}(x,x')+G_s^{\rm R}(x,-x')\sigma_2\Bigl),
\end{multline}
where $G_s^{\rm R}$ is GF in the extended space:
\be\label{eq:full_GF}
 G^{\rm R}_s(x,x') = \<x|\frac1{\ve-H_s^{\rm R}}|x'\>
\ee
To evaluate this expression, 
we will use the fact that a square of the $2\times 2$ Dirac Hamiltonian gives the 
Schr\"{o}dinger Hamiltonian.
First, we rewrite Eq.$\,$\eqref{eq:full_GF} multiplying both the numerator and denominator by  $(\ve+ H^R_s)$:
\be\label{eq:dominator_trick}
 G^{\rm R}_s(x,x') = \<x|(\ve+ H^R_s)\frac1{\ve^2- ({H_s^R})^2}|x'\>
\ee
Now the 
second term has the form of GF for 1D parabolic Schroedinger equation with delta function potential:
\be
({H_s^R})^2 = -v^2\partial_x^2+k^2v^2+\Delta^2-2v\mu_{k,s}\sigma_2\delta(x) = H_0+V(x)
\ee
where $H_0 = -v^2\partial_x^2+k^2v^2+\Delta^2$, and $V(x) = -2v\mu_{k,s}\sigma_2\delta(x)$.

The second factor in Eq.$\,$\eqref{eq:dominator_trick} can be represented as geometric series
\be
\frac1{\ve^2- ({H_s^R})^2} = G_0+G_0VG_0+\dots,\qquad G_0 = \frac 1{\ve-H_0}
\ee 
These series can be 
evaluated using the T-matrix formalism:
\be\label{eq:t-matrix_trick}
\frac1{\ve^2-({H_s^R})^2} = D_k+D_k|0\> T_s\<0| D_k,
\ee
where
\be
T_s = -\frac{2v\mu_{k,s}\sigma_2}{1+2v\mu_{k,s}\sigma_2\<0|D_k|0\>},\qquad D_k = \frac 1{\ve^2-h_k^2}.
\ee
Combining Eq.$\,$\eqref{eq:dominator_trick} and Eq.$\,$\eqref{eq:t-matrix_trick}, we 
obtain 
\begin{multline}
G^R_s(x,x') = \ve D_k(x,x')+\<x|H^R_sD_k|x'\>+\\
+\Bigl(\ve D_k(x,0)+\<x|H^R_s D_k|0\>\Bigl)D_k(0,x') T_s
\end{multline}
The quantities in this sum 
can be obtained by inverse Fourier transform\\
\begin{align*}
&D_k(x,x') = \<x|D_k|x'\> = -\int\limits_{-\infty}^{\infty}\frac{dp}{2\pi} \frac {{\rm e}^{-ip(x-x')}}{v^2p^2+m_k^2} 
= -\frac{e^{-m_k|x-x'|/v}}{2m_k v} 
,\nonumber\\
&\<x| H^R_sD_k|x'\> = -\int_{-\infty}^{\infty}\frac{dp}{2\pi} \Bigl(pv\s_1+\epsilon_{k,s}\s_2+ \mu_{k,s}\s_3{\rm sign}(x)\Bigl)
\\ \nonumber
& \times\frac {{\rm e}^{-ip(x-x')}}{v^2p^2+m_k^2}=-\frac{e^{-m_k|x-x'|/v}}{2m_k v}
\\\nonumber 
 &\times \Bigl(im_k v \sigma_1\,\text{sign}(x-x')+\ve_{k,s}\sigma_2+\mu_{k,s}\sigma_3{\rm sign}(x)\Bigl),
\end{align*}
where 
we defined $m_k = \sqrt{v^2k^2+\Delta^2-\ve^2}$. The quantity $m_k$, which may be either real or imaginary, represents an off-shell value of the parameter $\mu_k$ introduced above to describe wavefunction decay in the bulk.

Using these results we first 
derive the expression  for 
the full-space GF $G^R_s(x,x')$.  
Then, using Eq.$\,$\eqref{eq:mirror_trick}, we calculate the equal-point GF in the physical halfspace $G_s(x'=x)$, $x>0$.  Taking trace and imaginary part we finally obtain the spatially resolved DOS $\nu(\ve,x)$:
\begin{widetext}
\begin{align} \label{eq:full_greens_function}
&G^R_s(x,x') = -\frac1{2m_k v}\Bigl(\ve+im_k v \sigma_1\text{sign}\Delta x+\epsilon_{k,s}\sigma_2+\mu_{k,s}\sigma_3\text{sign}(x)\Bigl) \biggl( e^{-m_k|\Delta x|/v}+\frac{\mu_{k,s}\sigma_2}{m_k-\mu_{k,s}\sigma_2}e^{-m_k\bigl(|x|+|x'|\bigl)/v} \biggl)
,\\ \label{eq:reduced_greens_function}
&G_s(x'=x) = \lim_{x\rightarrow x'}G_s(x,x') =
-\frac 1{4m_k v}\Bigl(\ve+\epsilon_{k,s}\sigma_2+\mu_{k,s}\sigma_3{\rm sign}(x)\Bigl)\biggl(1+\frac{m_k\sigma_2+\mu_{k,s}}{m_k-\mu_{k,s}}\exp\bigl(-2m_k x/v\bigl)\biggl)
,
\\
&\nu(\ve,x) = \frac 1{\pi v} \Im\sum_k\frac1{m_k} \Biggl(\ve+\frac 12\sum_{s=\pm 1}\frac{\ve_{k,s}m_k+\ve\mu_{k,s}}{m_k-\mu_{k,s}}\exp(-2m_k x/v)\Biggl)
,
\end{align}
\end{widetext}
where $\Delta x = x-x'$.

The sum over in-plane surface momentum $k$ in the formula for $\nu(\ve,x)$ can be divided into two parts. 
The function under the sum has a continuous imaginary part for $|k|<v^{-1}\sqrt{\ve^2-\Delta^2}$, which 
defines the contribution of the bulk states $\nu_{\rm bulk}(\ve,x)$. For $|k|> v^{-1}\sqrt{\ve^2-\Delta^2}$ the function under the sum is real and continuous almost everywhere except of poles given by the condition $m_k-\mu_{k,s}=0$. Contributions from the poles into the sum represent the density of surface states $\nu_{\rm surf}(\ve,x)$. In these notations,
\be \label{eq:full_dos}
\nu(\ve,x)=\nu_{\rm bulk}(\ve,x)+\nu_{\rm surf}(\ve,x)
\ee
The first term can be 
split into a sum of DOS $\nu_0(\ve)$ for the homogeneous material 
and 
a boundary contribution that vanishes asymptotically in the bulk. For $d$-dimensional material:
\be\label{eq:dos_bulk}
\begin{split}
&\nu_{\rm bulk}(\ve,x) = \nu_0(\ve)+\frac{1}{2\pi v}\Im\sum_{s=\pm1}\int_{-k_\ve}^{k_\ve}\frac{d^{d-1}k}{(2\pi)^{d-1}}\frac{1}{m_k}
\\ 
&\times \frac{\ve_{k,s}m_k+\ve\mu_{k,s}}{m_k-\mu_{k,s}}\exp(-2m_k x/v),\quad k_\ve = \sqrt{\ve^2-\Delta^2}/v
\end{split}
\ee
For graphene the bulk DOS equals 
\be
\nu_0(\ve) =\begin{cases}
\frac{|\ve|}{2\pi v^2},\quad &|\ve|>\Delta\\
0, \quad &|\ve|<\Delta
\end{cases}
\ee
which corresponds to a gapped graphene band. 
For a 3D Dirac material,
\be
\nu_0(\ve) = \begin{cases}\frac{|\ve|\sqrt{\ve^2-\Delta^2}}{2\pi^2 v^3},\quad &|\ve|>\Delta\\
0, \quad &|\ve|<\Delta
\end{cases}
\ee
In the 
second term in Eq.$\,$\eqref{eq:dos_bulk} the integral over momenta $k$ 
gives rise to Friedel oscillations near the surface. 
The function under the integral is analytic and thus we can evaluate the integral numerically. The poles at $\mu_k=m_k$ lie outside the integration domain $-k_\ve<k<k_\ve$ and do not present any problem.

In contrast, the 
surface states contribution $\nu_{\rm surf}(\ve,x)$ (second term of Eq.$\,$\eqref{eq:full_dos}) can be found in a closed form analytically. To evaluate this contribution we use the complex analysis to express the residues at the poles corresponding to the discrete momentum values $k(\ve)=\pm k_0$ defined by
\be
\quad m_{k_0}-\mu_{k_0,s} = 0.
\ee
The surface states contribution to overall DOS is
\begin{align*}\label{eq:dos_surf}
&\nu_{\rm surf}(\ve,x) = \frac{\Omega_{d-1} k_0^{d-2}}{(2\pi)^{d-1} m_{k_0}v}\frac{\ve \mu_{k_0,s}+m_{k_0}\ve_{k_0,s}}{\frac\partial{\partial k}(m_k-\mu_{k,s})\bigl|_{k=k_0}}e^{-2m_{k_0} x/v}
\\
&=\frac{\Omega_{d-1}}{(2\pi v\sin\theta_n)^{d-1}v}\frac{(\ve-\ve_0)\mu_0}{\bigl((\ve-\ve_0)^2-\Delta_0^2\bigl)^{\frac{3-d}2}}e^{-2\mu_0 x/v}
\end{align*}
where $\Omega_d$ is d-spherical volume element, $\ve_0 = \Delta\cos\theta_\tau\cos\theta_n$ is the surface states neutrality point, $\Delta_0=\Delta\sin\theta_\tau \sin\theta_n$ is the surface states gap, and $\mu_0 = (\Delta\cos\theta_\tau - \ve\cos\theta_n)/\sin\theta_n$ characterizes inverse confinement length of the surface states. The solution exists only for energies $\ve$ satisfying both $\mu_0>0$ and $|\ve-\ve_0|>\Delta_0$. After integration over $x$ coordinate, the expression for density coincides with DOS of $d-1$ dimensional free relativistic particle.

The spatially resolved DOS for 2D material with $\Delta=0$ (e.g. monolayer graphene) and $\theta_n=-\pi/4$ is plotted 
in Fig.S1. In Fig. 2D of the main text the same quantity is shown with the bulk contribution $\nu_0(\epsilon)$ subtracted. This helps to discern Friedel oscillations and amplify the contribution of surface states.  Fig.S1 illustrates that the 
the contribution of surface states to spatially resolved DOS 
is concentrated in a narrow region near the surface $x\lesssim  \ve\sin^{d-1}\theta_n/\hbar v$. In this region the contribution of surface states overwhelms all other contributions to DOS including $\nu_0(\epsilon)$.

\begin{figure}[t!]

\includegraphics[width=0.9\columnwidth]{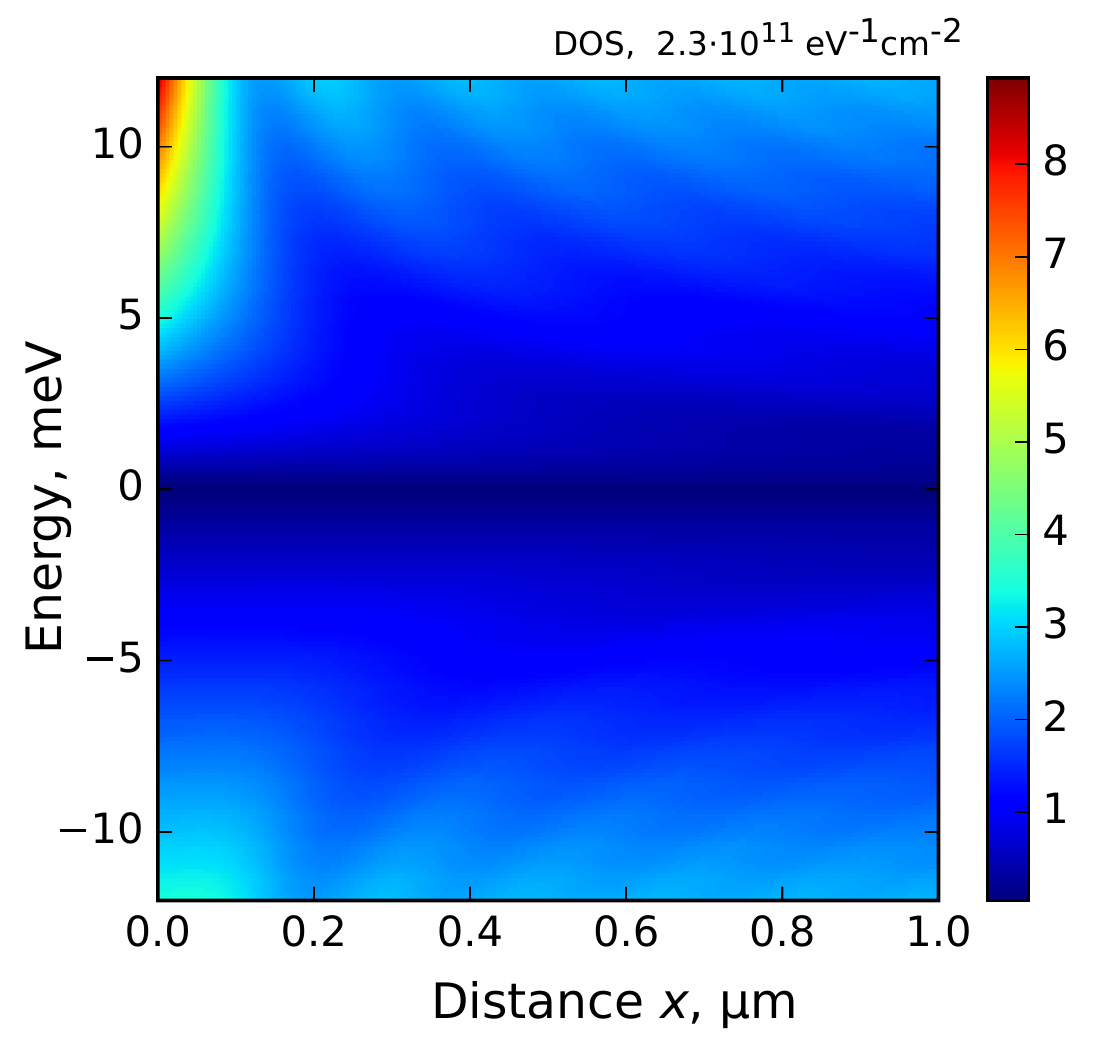}
\caption{The density of states (DOS) as a function of energy  
and distance from edge $\theta_\tau = 0$, $\theta_n = -\pi/4$ for monolayer graphene $d=2$, $\Delta=0$, and $v=10^6$ m/s as given by overall DOS expression in Eq.$\,$\eqref{eq:full_dos}.
Similar to Fig. 2D, the bright peak in DOS near the edge $x=0$ at positive energies describes the surface states contribution. }
\end{figure}

\section{Disorder at the edge} 

The surface of materials are typically highly disordered due to an imperfect lattice termination. To estimate the influence of disorder on surface transport, we are looking for the scattering rate and related mean free path of surface states.
A simple way to model the the surface disorder is introducing a non-correlated random potential $\xi({\bf r}_\parallel)$ at the surface:
\be \label{eq:random_potential}
V(x)\rightarrow V(x)+\xi({\bf r}_\parallel)\delta(x), \qquad \<\xi({\bf r}_\parallel)\xi({\bf r}'_\parallel)\>_{\rm dis}=\alpha\delta({\bf r}_\parallel-{\bf r'}_\parallel)
\ee
where ${\bf r}_\parallel$ is coordinate vector components parallel to the surface, $\delta(x)$ is the Dirac delta function and $\alpha$ is potential strength parameter (in $d$-dimensional material $\alpha$ has a dimensionality of energy$^2\times$space$^{d+1}$).

  The scattering rate can be found from the imaginary part of poles of disorder-averaged Green's function (GF) $\tilde G_s$. For a weak disorder, the GF can be represented as a perturbative series using the disorder potential strength as a small parameter:
\be
\tilde G_s = G_s+G_s\delta VG_s+G_s\delta VG_s\delta VG_s+\dots
\ee
where $G_s$ is a clean system GF obtained in the previous section (see Eq.$\,$\eqref{eq:system_GF} and Eq.$\,$\eqref{eq:reduced_greens_function}).

In the non-crossing approximation, the disorder-averaged GF can be expressed in terms of self energy $\Sigma$:
\be
\<\tilde G_s\>_\text{dis} =G_s+\alpha G_s\Sigma G_s+\alpha^2G_s\Sigma G_s\Sigma G_s+\dots
\ee
where
\be
 \Sigma = |0\>\<0|\sum_k g_{k,s} \qquad g_{k,s}= G_s(0)
\ee
Here $|0\>$ is a state with coordinate $x=0$ and $G_s$ depends on momentum $k$ being a parameter in the Hamiltonian (see Eq.$\,$\eqref{eq:reduced_greens_function}).

In the lowest order of perturbation theory using small parameter $\alpha$, the poles of the disorder-averaged Green's functions can be represented as function of spatially resolved DOS at the surface $\nu(\ve,x)$ (see previous section):
\begin{multline}
\tilde \ve_{k,s} = \ve_{k,s}+2\alpha\frac{\mu_{k,s}}{v}\Tr\sum_k \<0|G_s|0\>+O(\alpha^2) = \\
\ve_{k,s}-2\pi \alpha\frac{\mu_{k,s}}v \nu(\ve_{k,s},0) +O(\alpha^2)
\end{multline}
Imaginary part of the poles define the scattering time $\tau$:
\be
-\frac1{2\tau} = \Im \tilde \ve_{k,s} = -2 \pi\alpha\frac{\mu_{k,s}}v \nu(\ve_{k,s},0)
\ee
The mean free path is proportional to the scattering time and group velocity of surface states $u_{k,s}$: 
\be\label{eq:l_general}
\ell = u_{k,s}\tau =  \frac {vu_{k,s}}{4\pi\alpha\mu_{k,s} \nu(\ve_k,0)}, \qquad u_{k,s}=\left|\frac{\partial\ve_{k,s}}{\partial k}\right|,
\ee
At the surface layer, the largest contribution to DOS is given by surface states density. Therefore we neglect the bulk states DOS in the calculation of mean free path. For 2D materials with arbitrary boundary conditions, the DOS contribution is
\be
\nu(\ve_{k,s},0) = \frac{\Delta\cos\theta_\tau-\ve_{k,s}\cos\theta_n}{2\pi v^2\sin^2\theta_n}\left(1-\frac{\Delta_0^2}{(\ve_{k,s}-\ve_0)^2}\right)^{-1/2},
\ee
and the group velocity is
\be
u_{k,s} = v\sin\theta_n\left(1-\frac{\Delta_0^2}{(\ve_{k,s}-\ve_0)^2}\right)^{1/2},
\ee
where $\ve_0 = \Delta \cos\theta_\tau\cos\theta_n$, and $\Delta_0=\Delta\sin\theta_\tau \sin\theta_n$.

This allows us to express the mean free path as a function of the phase parameter $\theta_n$:
\be\label{eq:mfp_general}
\ell = \frac{v^4\sin^4\theta_n}{2\alpha(\ve-\Delta\cos\theta_\tau/\cos\theta_n)^2\cos^2\theta_n}\left(1-\frac{\Delta_0^2}{(\ve-\ve_0)^2}\right)
\ee

For gapless 2D materials $\Delta=0$ (e.g. monolayer graphene) and zigzag type of boundary, $\theta_n = \theta_V$, $\theta_\tau = 0$:
\be\label{eq:mean_free_path}
\ell = \frac{v^4\sin^4\theta_V}{2\alpha\ve^2\cos^2\theta_V} = \lambda^2 \frac{v^2\sin^2\theta_V}{8\pi^2\alpha\cos^2\theta_V}
\ee
where $\lambda = 2\pi/k$ is surface states wavelength. The expression Eq.$\,$\eqref{eq:mean_free_path} immediately leads to the Eq.(18) in the main text.

As it was shown is the main text, in realistic sample the mean free path can be much larger than relevant system size in experiments. Formula Eq.$\,$\eqref{eq:mfp_general} suggests that exclusions are possible if the parameter $\theta_n$ is small or the Fermi level is close to the band top/bottom $\ve\rightarrow\ve_0\pm\Delta_0$.

\section{Surface states in magnetic field} 

In this section we study the effect of magnetic field on Dirac surface states in 3D materials. The main motivation for this problem is the difference in charge carriers behavior in systems of two and three space dimensions. In a pure 2D system, charge carriers form a series of highly degenerate Landau levels (LL). In 3D systems this degeneracy is essentially broken and  LL become broadened due to presence of the momentum in the direction of the magnetic field. 

In previous sections we demonstrated that in a number of situations surface states behave similarly to free 2D modes. Thus, if the magnetic field is applied normally to the surface, one can expect that the spectrum of surface states should also be discrete due to absence of normal momentum broadening, as for 2D charge carriers. To study this effect, we use a simple model of electrons confined by 2D delta scalar potential.

We choose the shape delta of potential coinciding with $yz$ plane crossing $x$ axis at the point $x=0$. We choose the direction of magnetic field to be along $x$ axis such that $ \vec{B} = (B,0,0)$, $B>0$. The strength of delta potential can be expressed through the phase parameter $\theta_V$ in Eq. (16) in the main text leading to the Hamiltonian in the form
\begin{equation}\label{eq:delta_function_ham}
H = \sigma_1 vp_x+\sigma_2 v\left(p_y+eBz\right)+\tau_2\sigma_3vp_z+\tau_3\sigma_3\Delta +v\theta_V\delta(x)
\end{equation}
where we used Maxwell units and the Landau gauge $\vec{A} = (0,-Bz,0)$. 

The object of our study is confined states, therefore we apply the following ansatz for the eigenstates:
\be\label{eq:delta_function_conditions}
\psi(\vec{r}) = 
\begin{cases}
\psi_-(\vec{r}) \qquad x>0 \\
\psi_+(\vec{r}) \qquad x<0 
\end{cases},\qquad \psi_\pm(\vec{r}) = {\rm e}^{ip_y y \pm \mu x}\phi_\pm(z)
\ee
where $\mu>0$ is a confinement rate, and two functions $\psi_\pm(\vec{r})$ are connected by the continuity relation
\begin{equation}\label{eq:continuity_relation}
\psi_+(\vec{r})\bigl|_{x\rightarrow 0_-} = \exp(i\theta_V \sigma_1)\psi_-(\vec{r})\bigl|_{x\rightarrow 0_+}
\end{equation}
Below we are looking for a general solution for $x<0$, while the general solution for $x>0$ can always be obtained by changing $\mu\rightarrow -\mu$.

First, we evaluate the solution in the absence of magnetic field, $B=0$. In this case we can ignore, without a loss of generality, the $z$ space dimension and rewrite the Hamiltonian in Eq.$\,$\eqref{eq:delta_function_ham} in the form
\begin{equation}
H = \sigma_1 vp_x+\sigma_2 v p_y+\tau_3\sigma_3\Delta +v\theta_V\delta(x)
\end{equation}
The solution for $x<0$ is a superposition of eigenstates of $\tau_3$ operator:
\begin{equation}\label{eq:free_solution}
\phi_- = \alpha\left(\begin{matrix}
iv(\mu-p_y) \\ \Delta-\ve \\0\\0 
\end{matrix}\right)+\beta\left(\begin{matrix}
0 \\ 0\\iv(\mu-p_y) \\ -\Delta-\ve 
\end{matrix}\right)  
\end{equation}
where $\ve^2 = \Delta^2+p_y^2-\mu^2v^2 $, $\alpha$ and $\beta$ are complex parameters.

The continuity conditions in Eq.$\,$\eqref{eq:continuity_relation} can be written separately  for the first and the second term of the superposition in the Eq.$\,$\eqref{eq:free_solution} leading to the equivalent conditions for existence of non-zero solutions:
\be\label{eq:mu_ve_relation}
\mu v\cos\theta_V+\ve\sin\theta_V = 0
\ee
The allowed energy values derived using Eq.$\,$\eqref{eq:mu_ve_relation} are
\be \label{eq:free_particle_sol}
\ve = -\sgn(\theta_V)\cos\theta_V\sqrt{\Delta^2+v^2p_y^2}
\ee
The spectrum is double degenerate, with the degeneracy described by $\tau_3$ projections $\tau_3=\pm1$. The expression coincides with the spectrum Eq.(8) in the main text derived for armchair edge $\theta_\tau = \pi/2$, $\theta_n = \pi/2$.

The solution for non-zero magnetic field $B$ can be derived is a similar fashion. After a shift $z\rightarrow z-p_y/eB$, the equation $(H-\ve)\psi=0$ is transformed to the form
\begin{equation}\label{eq:B_eq_set}
\begin{cases}
 (\Delta -\epsilon)\psi_1+ iv (\mu -eBz)\psi_2 -i vp_z\psi_3=0\\
 i v(eBz+ \mu)\psi_1  -(\Delta +\epsilon)\psi_2+ i v p_z\psi_4=0 \\
 i vp_z\psi_1 -(\Delta +\epsilon)\psi_3  + i v(\mu -eBz)\psi_4 =0\\
 -i vp_z\psi_2 + i v(eBz+ \mu)\psi_3  +(\Delta -\epsilon)\psi_4=0
\end{cases}
\end{equation}
This set of equation can be rewritten in the form of four codependent quantum harmonic oscillator equations: 
\begin{equation}\label{eq:harmonic_osc_set}
\begin{cases}
 (\ve^2-\Delta^2+\mu^2\pm1)(i\psi_1\pm \psi_4) =H_0(i\psi_1\pm \psi_4) \\
 (\ve^2-\Delta^2+\mu^2\pm1)(i\psi_2\pm \psi_3) =H_0(i\psi_2\pm \psi_3)
\end{cases}
\end{equation}
where
\be
H_0 = \frac12 p_z^2+\frac {(veB)^2}2 z^2
\ee
The solution for the harmonic oscillator is integrable and has a discrete spectrum of eigenvalues:
\be
H_0f_n(z) = veB\Bigl(n+\frac12\Bigl)f_n(z),\qquad n\geq0
\ee 
where $f_{n}(z)\sim \exp(-veBz^2/2)H_n(\sqrt{veB}z)$ are eigenstates, $H_n$ are Hermite polynomials. Comparing it to Eq.$\,$\eqref{eq:harmonic_osc_set}, we get the values for system energies:
\begin{equation}\label{eq:dispersion}
\ve_n^2 = 2veBn+\Delta^2-v^2\mu_n^2, \qquad n\geq 0
\end{equation}
and the most general form of eigenstate components
\begin{equation}
\begin{matrix}
\psi^n_1 = \alpha f_n(z)+\beta f_{n-1}(z)\\
\psi^n_4 = i\Bigl(\alpha f_n(z)-\beta f_{n-1}(z)\Bigl)\end{matrix},\quad
\begin{matrix}
\psi^n_2 = \alpha' f_n(z)+\beta' f_{n-1}(z)\\
\psi^n_3 = i\Bigl(\alpha' f_n(z)-\beta' f_{n-1}(z)\Bigl)
\end{matrix}
\end{equation}
where $\alpha$, $\alpha'$, $\beta$, $\beta'$ are complex parameters, and by definition we set $f_{-1}(z)\equiv0$. Values of $\mu_n$ can be found by applying the continuity conditions in Eq.$\,$\eqref{eq:delta_function_conditions}.

For $n>0$, the $z$ dependent part of the eigenstates in Eq.$\,$\eqref{eq:delta_function_conditions} has the form:
\begin{equation}\label{eq:B_ss_solution}
\phi_-(z) = \left(\begin{matrix}
\alpha \\ \alpha'\\i\alpha'\\i\alpha 
\end{matrix}\right)f_n(z)+\left(\begin{matrix}
\beta \\ \beta'\\-i\beta'\\-i\beta 
\end{matrix}\right)f_{n-1}(z) 
\end{equation}
The dependence between parameters in this representation can be established after a substitution of Eq.$\,$\eqref{eq:B_ss_solution} into the original set of equations Eq.$\,$\eqref{eq:B_eq_set}:
\be
\begin{cases}
(\Delta-\ve_n)\alpha+iv\mu_n\alpha' = i\sqrt{2veBn}\beta'\\
iv\mu_n\alpha -(\Delta+\ve_n)\alpha' = -i\sqrt{2veBn}\beta
\end{cases}
\ee
Solving the system of equation relative $\alpha$ and $\alpha'$, we derive
\begin{align}
&\alpha = -\frac{v\mu_n\beta}{\sqrt{2veBn}}-\frac{i(\ve_n+\Delta)\beta'}{\sqrt{2veBn}},\\
&\alpha' = \frac{i(\ve_n-\Delta)\beta}{\sqrt{2veBn}}+\frac{v\mu_n\beta'}{\sqrt{2veBn}}
\end{align}
These values must be substituted into Eq.$\,$\eqref{eq:B_ss_solution}. The next step is resolving continuity relation in Eq.$\,$\eqref{eq:continuity_relation}. For real positive $\mu$ it has non-zeros solutions only if
%
\be\label{eq:nu_ve_relation}
v\mu_n\cos\theta_V+\ve_n\sin\theta_V = 0
\ee
Combining Eq.$\,$\eqref{eq:nu_ve_relation} and Eq.$\,$\eqref{eq:dispersion}, we get the allowed energy values:
\be\label{eq:spectrum_part1}
\ve = -{\rm sign}(\sin\theta_V)\cos\theta_V \sqrt{2veBn+\Delta^2}
\ee
For $n=0$ the the eigenstate Eq.$\,$\eqref{eq:delta_function_conditions} has only one  mode:
\begin{equation}
\phi_-(z) = \left(\begin{matrix}
\alpha \\ \alpha'\\i\alpha'\\i\alpha 
\end{matrix}\right)f_0(z),\qquad \alpha' = \frac{iv\mu_0}{\Delta+\ve_0}\alpha
\end{equation}

Similar to the case $n>0$, the continuity condition can be resolved only if the equality Eq.$\,$\eqref{eq:nu_ve_relation} holds. The only allowed energy value for $n=0$ is 
\be\label{eq:spectrum_part2}
\ve_0 = -{\rm sign}(\sin\theta_V)\cos\theta_V\Delta
\ee
Merging the solutions for $n>0$ in Eq.$\,$\eqref{eq:spectrum_part1} and $n=0$ inEq.$\,$\eqref{eq:spectrum_part2}, we obtain the Landau levels for the system:
\be\label{eq:ss_spectrum_B}
\ve_n = -{\rm sign}(\sin\theta_V)\cos\theta_V \sqrt{2veBn+\Delta^2},\qquad n=0,1,2\dots
\ee
The solution exists only for one sign of energy, positive or negative. The sign of the allowed energies coincides with the sign of $B=0$ solutions Eq.$\,$\eqref{eq:free_particle_sol}. This essential particle hole asymmetry can be used in experiment to distinguish the surface states from bulk modes in 3D Dirac materials.\\

\onecolumngrid
\vspace{1cm}
\begin{small}
\noindent[1] Davison, Sydney G., and Maria Stęślicka. Vol. 46. Oxford University Press, 1992.
\vspace{0.1cm}

\noindent[2] Lifshitz, I. M., and S. I. Pekar. Phys.-Usp 56.4 (1955).
\vspace{0.1cm}

\noindent[3] Kronig, R. de L., and W. G. Penney Proc. R. Soc. Lond. A 130.814 (1931): 499-513.
\vspace{0.1cm}

\noindent[4] Flugge, Siegfried. Practical quantum mechanics. Springer Science \& Business Media, 2012.
\end{small}
\end{document}